\documentstyle{article}
\newcommand{\Bbb}[1]{{\bf #1}}
\newcommand{\frak}[1]{{\bf #1}}
\newcommand{\Z}{{\Bbb Z}}

\newcommand{\C}{{\Bbb C}}

\newcommand{\Ref}[1]{{\rm{(}\ref{#1}\rm{)}}}

\newcommand{\ben}{\begin{equation}}
\newcommand{\een}{\end{equation}}
\newcommand{\bean}{\begin{eqnarray}}
\newcommand{\eean}{\end{eqnarray}}
\newcommand{\be}{\begin{displaymath}}
\newcommand{\ee}{\end{displaymath}}
\newcommand{\bea}{\begin{eqnarray*}}
\newcommand{\eea}{\end{eqnarray*}}
\newcommand{\g}{{{\frak g}\,}}

\newcommand{\h}{{{\frak h\,}}}

\newcommand{\Id}{{\rm Id}}

\newenvironment{definition}{\par\vspace{.5\baselineskip}
\noindent{\it Definition\/}:}{\par\vspace{.5\baselineskip}}
\newtheorem%
{thm}{Theorem}
\newtheorem%
{proposition}[thm]{Proposition}
\newtheorem%
{lemma}[thm]{Lemma}
\newtheorem%
{corollary}[thm]{Corollary}
\newcommand{\End}{{\rm{End}}}
\newcommand{\Hom}{{\rm Hom}}
\newcommand{\RR}[3]{R^{(#1#2)}(z_{#1#2},\lambda#3)}
\author{Giovanni Felder\\
Department of Mathematics,\\
University of North Carolina at Chapel Hill,\\
Chapel Hill, NC 27599-3250, USA%
}
\title{Elliptic quantum groups\thanks{Supported in part
by NSF grant DMS-9400841. To appear in the Proceedings
of the International Congress of Mathematical Physics, Paris 1994}}
\date{December 1994}
\begin{document}
\maketitle

\noindent This note gives an account of a construction of
an ``elliptic quantum group'' associated with each
simple classical Lie algebra. It is closely related
to elliptic face models of statistical mechanics, and,
in its semiclassical limit, to the Wess-Zumino-Witten
model of conformal field theory on tori. More details
are presented in \cite{Fe} and complete proofs will appear in a separate
publication.

Quantum groups (Drinfeld-Jimbo quantum enveloping algebras, Yangians,
Sklyanin algebras, see \cite{D}, \cite{Sk}) are the algebraic
structures underlying integrable models of statistical mechanics and
2-dimensional conformal field theory, and found applications in
several other contexts. However, from the point of view of statistical
mechanics, the picture is not quite complete. In particular, elliptic
interaction-round-a-face models of statistical mechanics have sofar
escaped a description in terms of quantum groups (expect in the $sl_N$
case). In this paper, we give such a description. It is hoped that the
construction will shed light in other contexts, such as a description
of the category of representation of quantum affine Kac--Moody
algebras, or the elliptic version of Macdonald's theory.

Our definition is motivated by the following
known construction that links conformal field
theory to the semiclassical version of quantum
groups. Conformal blocks of WZW conformal
field theory on the
plane obey the consistent system of Knizhnik-Zamolodchikov (KZ)
differential equations for a function $u(z_1,\dots,z_n)$ taking
values in the tensor product of $n$ finite dimensional
representations of a simple Lie algebra $\g$ \cite{KZ}:
\ben\label{KZ}
\partial_{z_i}u=\sum_{j:j\neq i}r(z_i-z_j)^{(ij)}u
\een
Here, the ``classical $r$-matrix'' $r(z)$ is the tensor
$C/z$, where $C\in\g\otimes\g$ is a symmetric invariant
tensor. We use the notation $X^{(i)}$, for $X\in\g$
or End$(V_i)$, to denote
 $\Id\otimes\cdots\Id\otimes X\otimes\Id
\otimes\cdots\otimes\Id$, an element of $U(\g)^{\otimes n}$
(or an endomorphism of the tensor
product $\otimes_jV_j$).
Similarly, if $X=\Sigma_\alpha x_\alpha\otimes y_\alpha$,
$X^{(ij)}$ means $\Sigma_\alpha x^{(i)}_\alpha y^{(j)}_\alpha$.

The classical Yang--Baxter equation in $\g^{\otimes 3}\subset
U(\g)^{\otimes 3}$
\ben\label{CYBE}
[r^{(12)}(z),r^{(13)}(z+w)+r^{(23)}(w)]+[r^{(13)}(z+w),
r^{(23)}(w)]=0,
\een
is, in this setting, the consistency condition of \Ref{KZ}.
More precisely, if $r(z)\in\g\otimes\g$ is a tensor
satisfying $r(z)+r^{(21)}(-z)=0$, then the system \Ref{KZ}
is consistent for all sets of representations if and only
if $r$ obeys the classical Yang--Baxter equation.
Therefore we can consider a Knizhnik--Zamolodchikov equation
for each solution of the classical Yang--Baxter equation.
Solutions were partially classified (with a non-degeneracy
hypothesis) in \cite{BD}, and come in three families,
rational, trigonometric and elliptic. It is however
only in the rational case that they have a direct relation
to conformal field theory.

The relation to integrable models of statistical mechanics
is based on the remark that \Ref{CYBE} in
$\End(V\otimes V\otimes V)$ for some vector
space $V$, is the semiclassical
limit of the  Yang--Baxter equation
\be\label{YBE}
R^{(12)}(z)
R^{(13)}(z+w)
R^{(23)}(w)
=
R^{(23)}(w)
R^{(13)}(z+w)
R^{(12)}(z),
\ee
for a meromorphic map $\C\ni z\mapsto R(z)\in\End(V\otimes V)$.

Indeed, if we have a one parameter family of solutions
of the Yang--Baxter equation
with $R(z)=\Id-2\eta r(z)+O(\eta^2)$, as the parameter $\eta$
goes to zero,
then $r$ obeys the classical Yang--Baxter equation.
If $R$ is ``unitary'', i.e., if $R(z)R^{(21)}(-z)=\Id_{V\otimes V}$,
then $r(z)+r^{(21)}(-z)=0$.

Roughly speaking, to each solution of  the Yang--Baxter equation
there corresponds a bialgebra or quantum group, defined
by quadratic relations \cite{STF}. Starting from rational and
trigonometric  of the classical
Yang-Baxter equation, we arrive in this way to Yangians
and affine quantum universal enveloping algebras, which are Hopf
algebras, see \cite{D}. In the elliptic case, this construction
works only in the $sl_N$ case and leads to Sklyanin
algebras \cite{Sk}, \cite{Ch}. Also, the KZ equation
quantizes to a difference equation \cite{S}, \cite{FR}, which
in the rational and trigonometric case is an equation
for form factors of integrable quantum field theory in
two dimensions.

Let us see how the above construction can be generalized to
the genus one case.
Our starting point is the set of genus one
Kni\-zhnik--Za\-mo\-lo\-dchi\-kov--Ber\-nard
(KZB) equations,
obtained by Bernard \cite{Be1,Be2}
as generalization of the KZ equations. These equations have
been studied recently in
\cite{FaGa}, \cite{EK}, \cite{FW}, \cite{I}.

Let $\g$ be a simple complex
Lie algebra with invariant bilinear
form normalized in such a way that long roots
have square length 2. Fix a Cartan subalgebra $\h$.
 The KZB equations are equations
for a  function $u(z_1,\dots,z_n,\tau,\lambda)$ with values
in the weight zero subspace
(the subspace killed by $\h$)
of a tensor product of irreducible finite dimensional
 representations of $\g$. The arguments
$z_1,\dots z_n,\tau$ are complex numbers with $\tau$ in the
upper half plane, and the $z_i$ are distinct modulo the
lattice $\Z +\tau\Z$, and $\lambda\in\h$. Let us introduce
coordinates $\lambda=\Sigma\lambda_\nu h_\nu$ in terms
of an orthonormal basis ($h_\nu$) of $\h$.
In the formulation of \cite{FW}, the KZB equations
 take the form
\begin{equation}\label{KZB}
\kappa\partial_{z_j}u=
-\sum_\nu h_\nu^{(j)}\partial_{\lambda_\nu}u
+\sum_{l:l\neq j}
\Omega^{(j,l)}(z_j-z_l,\tau,\lambda)u.
\end{equation}
Here $\kappa$ is an integer parameter which is large
enough depending on the representations in the tensor
product and $\Omega\in\g\otimes\g$ is a  tensor
preserving the weight zero subspace that we now
describe. Let $\g=\h+\sum_{\alpha\in\Delta}\g_\alpha$ be
the root decompostion of $\g$, and $C\in S^2\g$ be the
symmetric invariant tensor dual to the invariant bilinear
form on $\g$. Write $C=\sum_{\alpha\in\Delta\cup\{0\}}C_\alpha$,
where $C_0=\sum_\nu h_\nu\otimes h_\nu$ and $C_\alpha\in
\g_\alpha\otimes\g_{-\alpha}$.
 Let $\theta_1(t,\tau)$ be Jacobi's theta function
\be
\theta_1(t,\tau)=-\sum_{j=-\infty}^{\infty}
e^{\pi i(j+\frac12)^2\tau+2\pi i(j+\frac12)(t+\frac12)}.
\ee
and introduce functions $\rho$, $\sigma$:
\begin{eqnarray*}
\rho(t)&=&\partial_t\log\theta_1(t,\tau),\\
\sigma(w,t)&=&
\frac{\theta_1(w-t,\tau)\partial_t\theta_1(0,\tau)}
{\theta_1(w,\tau)\theta_1(t,\tau)}.
\end{eqnarray*}
The tensor $\Omega$ is given by
\be
\Omega(z,\tau,\lambda)=
\rho(z)C_0+\sum_{\alpha\in\Delta}\sigma(\alpha(\lambda),z)C_\alpha
\ee
As shown in \cite{FW}, the functions $u$ from conformal field
theory have a special dependence on the parameter $\lambda$.
For fixed $z,\tau$, the function $u$, as a function of $\lambda$,
 belongs to a finite dimensional
space of antiinvariant
theta function of level $\kappa$ (and obey certain
vanishing conditions). Therefore the right way
to look at these equations is to consider $u$ as a function
of $z_1,\dots, z_n,\tau$ taking values in a finite dimensional
space of functions of $\lambda$.

The tensor $\Omega$ has the skew-symmetry property
 $\Omega(z)+\Omega^{(21)}(-z)=0$,
and commutes with
$X^{(1)}+X^{({2})}$ for all $X\in\h$.
The compatibility condition of \Ref{KZB} is then the
{\em modified classical Yang--Baxter  equation}
 \cite{FW}
\begin{eqnarray}\label{MCYBE}
\sum_\nu
\partial_{\lambda_\nu}\Omega^{(1,2)}h_\nu^{(3)}+
\sum_\nu\partial_{\lambda_\nu}\Omega^{(2,3)}h_\nu^{(1)}+
\sum_\nu\partial_{\lambda_\nu}\Omega^{(3,1)}h_\nu^{(2)} & &
\nonumber
\\
-[\Omega^{(1,2)},\Omega^{(1,3)}]
-[\Omega^{(1,2)},\Omega^{(2,3)}]
-[\Omega^{(1,3)},\Omega^{(2,3)}] &=& 0
\end{eqnarray}
in $\g\otimes\g\otimes\g$.
In this equation, $\Omega^{(ij)}$ is taken at $(z_i-z_j,\tau,\lambda)$.

Let us turn to the question of finding the quantum version
of the modified classical Yang--Baxter equation, which is
the elliptic version of the Yang--Baxter equation.
To do this, let us take some distance from Lie algebras
and consider the following setting.

Let $\h$ be the complexification of a Euclidean space $\h_r$ and
extend the scalar product to a bilinear form on $\h$. View $\h$ a an
Abelian Lie algebra. We consider finite dimensional diagonalizable
$\h$-modules $V$. This means that we have a weight decomposition
$V=\oplus_{\mu\in\h}V[\mu]$ such that $\lambda\in\h$ acts as
$(\mu,\lambda)$ on $V[\mu]$.  Let $P_\mu\in\End(V)$ be the projection
onto $V[\mu]$.

It is convenient to introduce the following notation.
Suppose $V_1,\dots, V_n$ are finite dimensional diagonalizable
$\h$-modules. If $f(\lambda)$ is a meromorphic function on $\h$ with
values in $\otimes_iV_i=V_1\otimes\cdots\otimes V_n$
or $\End{(\otimes_iV_i)}$, and $\eta_i$ are
complex numbers, we define a function on $\h$
\be
 f(\lambda+\sum\eta_ih^{(i)})=\sum_{\mu_1,\dots,\mu_n}
\prod_{i=1}^nP_{\mu_i}^{(i)}f(\lambda+\Sigma\eta_i\mu_i),
\ee
taking values in the same space as $f$.

Given $\h$ and $V$ as above,
the quantization of \Ref{MCYBE} is  an equation
for a meromorphic function $R$ of the spectral parameter
$z\in\C$ and an additional variable $\lambda\in\h$,
taking values in $\End(V\otimes V)$
\begin{eqnarray}
 &\RR 12{+\eta h^{(3)}}
\RR 13{-\eta h^{(2)}}
\RR 23{+\eta h^{(1)}}=& \nonumber\\
 &=
\RR 23{-\eta h^{(1)}}
\RR 13{+\eta h^{(2)}}
\RR 12{-\eta h^{(3)}}.&
\label{MYBE}\end{eqnarray}
Here $\eta$ is a parameter
and $z_{ij}$ stands for $z_i-z_j$.
This equation forms the basis for the subsequent
analysis. Let us call it modified Yang--Baxter equation
(MYBE). Note that a similar equation,
without spectral parameter, has appeared for the monodromy
matrices in Liouville theory, see
\cite{GN}, \cite{Ba}, \cite{AF}.
We supplement it by the ``unitarity'' condition
\ben
\label{UC}
\RR 12{}\RR 21{}={\rm{Id}}_{V\otimes V},
\een
 and the ``weight zero''
condition
\ben\label{WZ}
[X^{(1)}+X^{(2)},R(z,\lambda)]=0,\qquad \forall X\in\h.
\een
We say that  $R\in\End(V\otimes V)$ is a {\em generalized quantum $R$-matrix}
if it obeys \Ref{MYBE}, \Ref{UC}, \Ref{WZ}.

If we have a family of solutions parametrized by $\eta$
(the {\em same} parameter entering the MYBE) in
some neighborhood of the origin, and $R(z,\lambda)=
{\rm Id}_{V\otimes V}-2\eta \Omega(z,\lambda) +O(\eta^2)$ has
a ``semiclassical asymptotic expansion'', then \Ref{MYBE}
reduces to the modified classical Yang--Baxter
equation \Ref{MCYBE}.

Here are examples of solutions. Take $\h$ to be the
Abelian Lie algebra of
diagonal $N$ by $N$ complex matrices, with bilinear
form Trace($AB$),  acting on $V=\C^N$. Denote by $E_{ij}$ the
$N$ by $N$ matrix with a one in the $i$th row and $j$th column
and zeroes everywhere else. Then we have

\begin{proposition}
The function
\be\label{sol}
R(z,\lambda)=\sum_{i}E_{ii}\otimes E_{ii}
+\sum_{i\neq j}
\frac{\sigma(\gamma,\lambda_{ij})}
{\sigma(\gamma,z)}
E_{ii}\otimes E_{jj}
+\sum_{i\neq j}
\frac{\sigma(\lambda_{ij},z)}
{\sigma(\gamma,z)}
E_{ij}\otimes E_{ji},
\ee
is a ``unitary'' weight zero
solution of the modified Yang--Baxter equation, i.e., it
is a generalized quantum $R$-matrix,
with
$\eta=\gamma/2$.
\end{proposition}

\noindent Following the Leningrad school (see \cite{STF}, \cite{F}),
one associates a
bi\-al\-ge\-bra with qua\-dra\-tic relations
to each solution of the Yang--Baxter equation.
In our case we have modify slightly the construction. Let us
consider an ``algebra'' $A(R)$ associated to a generalized
quantum $R$-matrix $R$, generated by meromorphic functions
on $\h$ and the matrix elements (in some
basis of $V$) of a matrix $L(u,\lambda)\in\End(V)$ with
non commutative entries, subject to the relations
\bea &
R^{(12)}(z_{12},\lambda+\eta h)
L^{(1)}(z_1,\lambda-\eta h^{(2)})
L^{(2)}(z_2,\lambda+\eta h^{(1)})=& \\
&=
L^{(2)}(z_2,\lambda-\eta h^{(1)})
L^{(1)}(z_1,\lambda+\eta h^{(2)})
R^{(12)}(z_{12},\lambda-\eta h).&
\eea
Instead of giving a more precise definition of this algebra, let us
define the more important notion (for our purposes) of representation
of $A(R)$. We define a tensor category of ``representations
of $A(R)$''.

\begin{definition} Let $R\in\End(V\otimes V)$
be a meromorphic unitary weight zero solution of the MYBE
(a generalized quantum $R$-matrix). A
representation of $A(R)$ is a diagonalizable $\h$-module $W$ together
with a meromorphic function $L(u,\lambda)$ (called $L$-operator)
on $\C\times\h$ with values
in $\End(V\otimes W)$ such that the identity
\bea
&R^{(12)}(z_{12},\lambda+\eta h^{(3)})
L^{(13)}(z_1,\lambda-\eta h^{(2)})
L^{(23)}(z_2,\lambda+\eta h^{(1)})=&
\\
&=
L^{(23)}(z_2,\lambda-\eta h^{(1)})
L^{(13)}(z_1,\lambda+\eta h^{(2)})
R^{(12)}(z_{12},\lambda-\eta h^{(3)})&
\eea
holds in $\End(V\otimes V\otimes W)$, and so that
$L$ is of weight zero:
\be
[X^{(1)}+X^{(2)},L(u,\lambda)]=0, \qquad \forall X\in\h.
\ee
\end{definition}
\noindent We have natural notions of homomorphisms of representations:
A homomorphism $\phi:(W,L)\to (W',L')$ is a linear
map $\phi(u,\lambda)\in\Hom(W,W')$  depending meromorphically
on $u,\lambda$, such that $L'(u,\lambda)\Id\otimes\phi(u,\lambda)
=\Id\otimes\phi(u,\lambda)L(u,\lambda)$.

\begin{thm}\label{t1} (Existence and coassociativity of
the coproduct) Let $(W,L)$ and $(W',L')$  be representations
of $A(R)$. Then $W\otimes W'$ with $\h$-module
structure $X(w\otimes w')=Xw\otimes w'+w\otimes Xw'$ and
$L$-operator
\be
L^{(12)}(z,\lambda+\eta h^{(3)})
L^{(13)}(z,\lambda-\eta h^{(2)})
\ee
is a representation of $A(R)$.
 Moreover, if we have three
representation $W$, $W'$, $W''$, then the representations
$(W\otimes W')\otimes W''$ and $W\otimes(W'\otimes W'')$ are
isomorphic (with the obvious isomorphism).
\end{thm}

\noindent Note also that if $L(z,\lambda)$ is an $L$-operator then
also $L(z-w,\lambda)$ for any complex number $w$.
Since the MYBE and the weight zero condition
mean that $(V,R)$ is a representation,
we may construct representations on $V^{\otimes n}=
V\otimes\cdots\otimes
V$ by iterating the construction of Theorem \ref{t1}.
The corresponding $L$ operator is the ``monodromy matrix''
with parameters $z_1,\dots,z_n$:
\be
\prod_{j=2}^{n+1}R^{(1j)}(z-z_j,\lambda-\eta
\Sigma_{1<i<j}h^{(i)}+\eta\Sigma_{j<i\leq n+1}h^{(i)}).
\ee
(the factors are ordered from left to right). Although
the construction is very reminiscent of the Quantum Inverse
Scattering Method \cite{F}, we cannot at this point
construct commuting transfer matrices by taking the trace of the
monodromy matrices. Instead, as we will see now, one has to pass to
(interaction-round-a-) face models.

In our setting, the relation between the generalized quantum $R$-matrix and
the Boltzmann weights $W$ of the corresponding
interaction-round-a-face (IRF)
model \cite{B} is very simple. Let $R\in\End(V\otimes V)$
be a generalized quantum $R$-matrix, and let $V[\mu]$ be the component
of weight $\mu\in \h^*$ of $V$, with projection
$E[\mu]:V\to V[\mu]$. Then for $a,b,c,d\in\h^*$, such
that  $b-a$, $c-b$, $d-a$ and $c-d$ occur in the weight
decomposition of $V$, define
a linear map
\be
W(a,b,c,d,z,\lambda):V[d-a]\otimes V[c-d]\to
V[c-b]\otimes V[b-a],
\ee
by the formula
\ben
W(a,b,c,d,z,\lambda)=E[c-b]\otimes E[b-a]
R(z,\lambda+\eta a+\eta c)|_{V[d-a]\otimes V[c-d]}.
\een
Note that $W(a+x,b+x,c+x,d+x,z,\lambda-2\eta x)$ is independent
of $x\in\h\simeq\h^*$. Set $W(a,b,c,d,z)=W(a,b,c,d,z,0)$.

\begin{thm}
If $R$ is a solution of the MYBE,
 then $W(a,b,c,d,z)$ obeys the Star-Triangle relation
\bean\label{STR}
&\sum_g
W(b,c,d,g,z_{12})^{(12)}
W(a,b,g,f,z_{13})^{(13)}
W(f,g,d,e,z_{23})^{(23)}&
\\
&=
\sum_g
W(a,b,c,g,z_{23})^{(23)}
W(g,c,d,e,z_{13})^{(13)}
W(a,g,e,f,z_{12})^{(12)},&
\eean
on $V[f-a]\otimes V[e-f]\otimes V[d-e]$.
\end{thm}
The familiar form of the Star-Triangle relation \cite{B},\cite{JMO}
is recovered when the spaces $V[\mu]$ are 1-dimensional. Upon
choice of a basis, the
Boltzmann weights $W(a,b,c,d,z)$ are then numbers.

For example, if $R$ is the solution of Prop.\ \ref{sol}, we obtain
the well-known $A^{(1)}_{n-1}$ solution (see \cite{JMO},
\cite{JKMO} and references
therein).

Moreover we also have a converse of this theorem,
which gives us many examples of solutions of
the modified Yang--Baxter equation.

\begin{thm} Let $V$ be a finite dimensional diagonalizable
$\h$-module, with weight decomposition $\oplus_{\mu\in \h}V[\mu]$.
Set $A=\{\mu\in\h|V[\mu]\neq 0\}$.
Suppose that for each $(a,b,c,d)\in \h^4$ such
that $d-a$, $c-d$, $c-b$, $b-a\in A$,
 $W(a,b,c,d,z)\in\Hom(V[d-a]\otimes V[c-d],
V[c-b]\otimes V[b-a])$ is a meromorphic function of
$z$, and that these functions obey the Star-Triangle
relation \Ref{STR}. Assume also
that $W$ obeys the relation
\be
\sum_g W(a,g,c,d,z)W(a,b,c,g,-z)=\delta_{bd}.
\ee
Then
\be
R(z,\lambda)=\sum_{a,b,c,d:a+c=\lambda/\eta}W(a,b,c,d,z)
E(d-a)\otimes E(c-d),
\ee
is a generalized quantum $R$-matrix.
\end{thm}

In particular, if we take the solutions of \cite{JMO},
which have $V$ as the vector representation of simple
Lie algebras of type $A,B,C,D$,
we obtain generalized
$R$-matrices and thus elliptic quantum groups associated
to all classical simple Lie algebras.

We have not discussed here the difference equations arising
as quantization of the Knizhnik--Zamolodchikov-Bernard
equations. See \cite{Fe} for some detail on this point.

\end{document}